# Objective and Subjective Probabilities in Quantum Mechanics

Leslie Ballentine

*Simon Fraser University, Burnaby, B.C. V5A1S6, Canada*

**Abstract.** The concept of probability was prominent in the original foundations of quantum mechanics, and continues to be so today. Indeed, the controversies regarding objective and subjective interpretations of probability have again become active. I argue that, although both objective and subjective probabilities have domains of relevance in QM, their roles are quite distinct. Even where both are legitimate, the objective and subjective probabilities differ, both conceptually and numerically. There are quantum probabilities that have no useful subjective interpretations, and there are subjective probabilities that cannot be realized as quantum probabilities.

**Keywords:** quantum probability, objective, subjective

**PACS:** 03.65.Ta, 03.67.-a, 02.50.Cw

## 1. INTRODUCTION

Probability has played an important role in the Foundations of Quantum Mechanics (FQM) from the beginning, and continues to play an important role today. So it is appropriate, at this conference devoted to a reconsideration of FQM, to examine the role, or roles, of probability in QM.

Here are some of the ways in which Probability and FQM are inter-related:

- QM is formulated as an intrinsically probabilistic theory, rather than as a deterministic theory in which probability would play a secondary role.
- The choice of an interpretation of probability affects the interpretation of QM.
- Recent developments in Quantum Information Theory have led to new ways to look at FQM, including a greater emphasis on possible roles for Subjective Probability in QM. [1]

These ideas motivate my paper, although I shall not treat all of them in detail.

## 2. INTERPRETATIONS OF PROBABILITY

At an earlier conference here in Växjö (25 Nov. - 1 Dec. 2000) [2], I presented, in the spirit of Linnaeus, a taxonomic classification of the many interpretations of probability [3]. These were divided into four major groups:

1. **The Logic of Inductive Inference**. Probability is assigned to *propositions*; $P(A|C)$ is the probability that $A$ is true, given the information $C$. Inductive logic reduces to deductive logic in the case that the probabilities take on only the values 0 and 1. This point of view was expressed most completely by E. T. Jaynes in his posthumous book *Probability Theory: The Logic of Science* [4].
2. **Ensemble and Frequency interpretations**. Probability is identified with the *measure* on a set, or with the *limit frequency* in an ordered sequence.
3. **Subjective Probability**. A measure of the *degree of reasonable belief* under the condition of incomplete knowledge.
4. **Propensity**. This is a form of causality that is weaker than determinism, and is associated with *events* in an (apparently) indeterministic system. $P(A|C)$ is the propensity for event $A$ to occur under the condition $C$.

In my 2000 talk, I further suggested that the first of these interpretations (Inductive Logic) might be the master interpretation of probability, from which the others could be derived as special cases.

> "If we specialize to propositions about repeated experiments we obtain the Ensemble–Frequency theory. If we specialize to propositions about personal belief we obtain Subjective probability. If we specialize to propositions about indeterministic or unpredictable events we obtain the Propensity theory." (Ref. [3], p. 72)

I stressed the tentative nature of my classification, and anticipated that some future refinements might be needed.

## 2.1. The place of Subjective Probability

At a later conference in Växjö (June 1–6, 2003), Rudiger Schack criticized the treatment of subjective probability in my classification.[1] He said that, contrary to my statement quoted above, what changes when we pass to the subjective interpretation is not the nature of the propositions, but what you say about them. The significance of this brief remark can best be illustrated by an example.

Consider a particular quantum state of a spin one–half particle,

$$|\psi\rangle = \frac{4}{5}|+\rangle + \frac{3}{5}|-\rangle \tag{1}$$

and the proposition

A: On the next measurement the spin will be "+".

From the formalism of quantum mechanics, we obtain the uninterpreted probability statement,

$$P(A|\psi) = 0.64 \tag{2}$$

Each interpretation of probability will give a different meaning to this statement.

Propensity –

The probability is 0.64 that the next measurement will yield "spin +".

Frequency –

In a long run of similar measurements on this state, the fraction of "spin +" will be (close to) 64%.

Subjective –

My degree of belief (or confidence) that the next measurement will yield "spin +" is 64%.

Note, however, the subjective interpretation does NOT say

There is a 64% probability that I believe the spin will be "+".

The proposition A is about the spin, not about my belief! Schack was correct in pointing out that the subjective interpretation does not deal with a restricted class of propositions (those which concern personal beliefs), but rather it affects what we say about all propositions, namely, our degree of belief in them.

It is, therefore, necessary to revise the classification of interpretations of probability [3] as it applies to the Subjective interpretation. Rather than being subordinate to the Inductive Logic theory and derived from it by restricting the class of propositions, the Subjective interpretation is parallel to it and can deal with the same class of propositions. The distinction arises from what is said about those propositions; an objective (but possibly imprecise) statement about the truth or falsity of the proposition in the first case, and a subjective statement about someones degree of belief in the second case. Both interpretations are based on the same formalism of Bayesian probability theory, and may reasonably be called the *objective* and *subjective* Bayesian interpretations. Indeed, Jaynes [4] switches back and forth between them effortlessly, and perhaps even unconsciously.

---

[1] Unfortunately, his critical remarks were not included in the conference proceedings.

## 3. QUANTUM PROBABILITY

Which of the various interpretation of probability are relevant to quantum mechanics? In my earlier paper [3], I showed how both the Frequency and Propensity interpretations are relevant. Regarding the Subjective interpretation, I said, "I have *never* seen a coherent exposition of QM based on a *subjective* interpretation of quantum probabilities as representing *knowledge*." This claim is likely to be controversial because the interpretation of probabilities as knowledge seems to be a tenet of the Copenhagen interpretation, to which at least lip service is still paid. In the rest of this paper, I shall expand upon that statement, with examples to illustrate the point.

First I must define what is meant by the term *quantum probability*. For the purposes of this paper, I shall use it to mean those probabilities that are calculated from the familiar QM formulas, such as:

- $|\psi(x)|^2$     The probability density for the particle to be found at position $x$.
- $|\langle a_k|\psi\rangle|^2$     The probability that an observable takes on the eigenvalue $a_k$.
- $Tr(\rho E_\omega)$     The probability that an observable lies within the subset $\omega$ of its spectrum. Here $\rho$ is the state operator and $E_\omega$ is a projector onto the relevant subset of the spectrum.

Other formulas are also possible, but all assume that a certain state (represented by $\psi$ or $\rho$) has been prepared, and the probability for an observable to take on some value is calculated. Within the Frequency or Propensity interpretations these quantum probabilities are *objective* in the sense that they do not depend on the observer or his[2] knowledge, but only on the physical situation described by the prepared state. I shall later consider whether they can also be given a *subjective* interpretation as representing someones knowledge.

## 4. OBJECTIVE AND SUBJECTIVE PROBABILITIES IN QM

Subjective probability is a natural tool to use in cases where different persons have different amounts of information, and hence are likely to make different predictions. Classical and quantum cryptography, whose goal is to share information with some people while denying it to others, are good examples. But the subjective probabilities that occur in quantum cryptography supervene at a high level of application, far above the fundamental quantum probabilities of the previous section. Quantum cryptography uses subjective probabilities because it is cryptography, not because it is quantum.

### 4.1. Objective and Subjective Density Matrices

To better see the roles of objective and subjective probabilities, we turn to a simple example. A single photon[3] is prepared in the vertical polarization state,

$$|\psi\rangle = |\updownarrow\rangle. \tag{3}$$

This state vector and the probabilities associated with it are *objective*; they describe the result of the physical operation of state preparation, and not anyone's knowledge of it.

We then tell Alice, who is unaware of the state preparation, that the polarization state is either vertical $|\updownarrow\rangle$ or horizontal $|\leftrightarrow\rangle$. We tell Bob that the polarization was prepared in one of four states: vertical $|\updownarrow\rangle$, horizontal $|\leftrightarrow\rangle$, +45 degrees $|\nearrow\rangle$, or −45 degrees $|\searrow\rangle$.[4]

Except for a possible phase factor, the vector $|\nearrow\rangle$ is the normalized sum of the vertical and horizontal states, and the vector $|\searrow\rangle$ is their normalized difference. Using the vertical and horizontal state vectors as a basis, Bob's four

---

[2] The word *observer* is a technical term in QM. It denotes an abstract entity having only a few attributes, including a frame of reference and the ability to measure an observable. It should not be thought of as a human person with attributes like consciousness or gender. An observer is traditionally referred to as "he" for the same reason that a ship is referred to as "she" — mere tradition. The hermaphroditic forms "he or she" and "he/she" should be deprecated.

[3] There is nothing special about a photon. Any object with a two dimensional state space would do just as well.

[4] The arrows for +45 and −45 should also be double-ended, but unfortunately Latex has no such symbols.

possible states can be represented by the following density matrices:

$$\rho_\updownarrow = \begin{pmatrix} 1 & 0 \\ 0 & 0 \end{pmatrix}, \quad \rho_\leftrightarrow = \begin{pmatrix} 0 & 0 \\ 0 & 1 \end{pmatrix}, \quad \rho_\nearrow = \begin{pmatrix} 1/2 & 1/2 \\ 1/2 & 1/2 \end{pmatrix}, \quad \rho_\searrow = \begin{pmatrix} 1/2 & -1/2 \\ -1/2 & 1/2 \end{pmatrix}. \tag{4}$$

Although the actual state of the photon is $\rho_\updownarrow$, Alice and Bob do not know this, and so they must use their incomplete information to make a subjective estimate of the density matrix. Alice knows that the polarization may be either vertical or horizontal, so she forms a mixed state from those two components, obtaining the *subjective* density matrix

$$\rho_S = \begin{pmatrix} 1/2 & 0 \\ 0 & 1/2 \end{pmatrix}. \tag{5}$$

Bob knows only that the polarization is in one of the four states (4), so he forms a mixed state from four components, and obtains the same subjective density matrix $\rho_S$. Notice that, although Alice and Bob have different information, they both arrive at the same subjective density matrix. Nevertheless, their future predictions will not necessarily be the same.

To show this, we perform a measurement that distinguishes between vertical and horizontal polarizations. The result, of course, confirms that the polarization is vertical. We, the readers, knew this in advance, because we knew that the prepared state was $\rho_\updownarrow$. But to Alice and Bob, the result of the measurement is new information, which they can use to obtain a better estimate of the photon state. Alice will now correctly infer that the initial state must have been $\rho_\updownarrow$, since the other possibility ($\rho_\leftrightarrow$) is ruled out by the new data. But Bob, who previously knew only that the initial state was one of four, can now only reduce the number of possibilities to three. However, he can use Bayes theorem to improve his subjective estimate.

### 4.2. Bayesian Inference of an Unknown State

Since the use of Bayes theorem in quantum theory may be unfamiliar to many readers, I shall give Bob's calculation in full detail. For our purposes, Bayes theorem [5] may be written as

$$P(X_j|Y_i \& C) = P(Y_i|X_j \& C) \frac{P(X_j|C)}{P(Y_i|C)} \tag{6}$$

where "&" denotes the logical "and". Here $X_j$ is one of the four possible initial states (4), and $Y_i$ is one of the possible results of the polarization measurement (either $\updownarrow$ or $\leftrightarrow$). $C$ denotes the background information that Bob possesses, including the rules of QM and the knowledge that the initial state was one of the four in (4), but excluding the result of the polarization measurement. Lacking any information that would prefer any of the four possible states, Bob assigns the *prior probability*[5] $P(X_j|C) = 1/4$. Similarly, before the measurement has been performed, the prior probability for the two possible results is $P(Y_i|C) = 1/2$.

The remaining term on the righthand side of eq.(6), $P(Y_i|X_j \& C)$, is just the probability of the measurement yielding a particular result for a given quantum state, and it can be calculated from the familiar formulas for quantum probability. The probability for obtaining the observed polarization ($\updownarrow$) in each of the possible initial states is, in a slightly modified, but obvious notation,

$$P(\updownarrow | |\updownarrow\rangle) = 1, \quad P(\updownarrow | |\leftrightarrow\rangle) = 0, \quad P(\updownarrow | |\nearrow\rangle) = 1/2, \quad P(\updownarrow | |\searrow\rangle) = 1/2. \tag{7}$$

Combining the three factors on the righthand side of (6), we obtain Bob's *posterior probability* for each of the four possible initial states, after the result of the polarization measurement has been taken into account. In an abbreviated notation, Bob's probabilities are

$$P(| |\updownarrow\rangle) = 1/2, \quad P(| |\leftrightarrow\rangle) = 0, \quad P(| |\nearrow\rangle) = 1/4, \quad P(| |\searrow\rangle) = 1/4. \tag{8}$$

This calculation confirms the earlier obvious conclusion that Bob will not be able to uniquely infer the initial prepared state of the photon, but it also quantifies his knowledge of what that initial state might have been.

---

[5] The correct term is "prior", meaning "prior to the result of the measurement", and not "a priori", which has a different, metaphysical meaning that is not relevant here.

### 4.3. Bob's new Subjective Density Matrix

Bob can use his posterior probabilities (8) to form a new mixture as his subjective density matrix. It will be

$$\rho_B = \frac{1}{2}\rho_\uparrow + \frac{1}{4}\rho_\nearrow + \frac{1}{4}\rho_\nwarrow = \begin{pmatrix} 3/4 & 0 \\ 0 & 1/4 \end{pmatrix}. \tag{9}$$

Several remarks are in order about the significance of this density matrix $\rho_B$.

- It is not an estimate of the final state of the photon after the measurement. Indeed, if the measurement was performed as a projective measurement (von Neumann type) then the final state will be $\rho_\uparrow$, and this would be known to Bob and Alice.
- It does not represent Bob's state of belief about the unknown (to him) initial state of the photon. Bob does not believe the initial state was $\rho_B$. Rather, he believes its was either $\rho_\uparrow$, $\rho_\nearrow$ or $\rho_\nwarrow$, with probabilities given by eq.(8).
- However, it is Bob's "safest estimate" of the unknown initial state, in the sense that predictions from it will have the lowest probable error. Even though he now knows that $\rho_\uparrow$ is the most probable value for the initial state, he should hedge his bets in case it turns out to be $\rho_\nearrow$ or $\rho_\nwarrow$. The use of $\rho_B$ for prediction accomplishes this "hedging".
- It is not appropriate to interpret $\rho_B$ as "a mixture of 75% vertical polarization and 25% horizontal polarization", because Bob knows that the horizontal polarization state has zero probability. Evidently, the subjective density matrix does not adequately represent Bob's state of knowledge.

## 5. CONCLUSIONS

Although only a simple example has been treated in this paper, it is sufficient to establish some broad conclusions. The first is that both *objective* and *subjective* states/probabilities may legitimately appear in the same problem. However, they are both conceptually and numerically distinct. The objective state operator (or density matrix) describes the result of the physical operation of state preparation. It is objective in the sense that it does not depend on any observer or on his knowledge.[6] The subjective states and probabilities describe the states of belief of various observers, whose knowledge of the objective situation is incomplete. The objective state of a physical system and the subjective state of someones knowledge are quite different things; neither can replace the other.

Quantum probabilities, as defined in Sec. 3, are naturally to be interpreted as objective if they are calculated from the *objective* state operator. But what if the same formulas are used with *subjective* density matrices? Could we then say the quantum probabilities can also have a subjective interpretation? There are serious difficulties with such a program. Firstly, we have shown that observers who have *different* states of knowledge may, nevertheless, assign the same subjective density matrix. Therefore, the density matrix is inadequate to describe a state of knowledge. Secondly, we have seen that two observers who share the same subjective density matrix may, nevertheless, make different predictions. So, it cannot be correct to assert that quantum states represent states of knowledge.

Another difference between objective quantum states and states of knowledge is illustrated in the mathematical fact that a mixed state operator can be represented as a mixture of pure states,

$$\rho = \sum_i w_i |\psi_i\rangle\langle\psi_i|, \tag{10}$$

in infinitely many ways [6, 7]. Now if $\rho$ represents an objective state (the result of a physical operation of state preparation), then the state operator $\rho$ is sufficient to make all possible predictions about future measurements. In this (and only this) sense, it can be said that the quantum state description is *complete*.[7] However, a subjective density matrix is not adequate to describe a state of knowledge. In the example studied, we saw that Bob's state of belief could be described by a particular mixture representation of his subjective density matrix $\rho_S$, but other mathematically valid representations of the form (10) did not correspond to his state of belief. In particular, the diagonal representation, which has some privilege mathematically, does not describe his state of belief. Therefore, if we attempt to use the formalism of quantum states and quantum probability to describe states of knowledge or belief, we find that the subjective quantum state operator is *not* a complete state description. This shows the unnaturalness of forcing a subjective interpretation onto quantum states.

---

[6] To describe the state as objective in this sense is not the same as claiming that the state vector or operator has ontic status as an "element of reality", a claim that is well known to be problematic.

[7] This does not conflict with the argument of Einstein, Podolsky and Rosen [8] that the quantum description of reality is *incomplete*, because they used a different criterion for *completeness*.

Most of my discussion has dealt with mixed states because the mixture coefficients, $w_i$, in (10) may sometimes be given a meaningful subjective interpretation. But what about the pure states, such as $\psi(x)$? Can pure state probabilities be given a meaningful interpretation as knowledge? What state of knowledge would lead someone to describe the probability density for the electron in the ground state of hydrogen by its particular analytic form? What state of knowledge would lead to the probability density for an electron in a covalent bond (a function that has no analytic form, but which can be accurately computed numerically)? These probabilities have a natural objective interpretation, being experimentally testable in reproducibly prepared states. But they have no natural interpretation as someones knowledge, except for the contrived case of a person who knows and believes everything QM tells him.

## ACKNOWLEDGMENTS

I have enjoyed conversations with Rudiger Schack and Chris Fuchs on this subject. We may still disagree, but at least we now better understand why.## REFERENCES

1. A. Khrennikov, editor, Proceedings of the Conference, *Quantum Theory: Reconsideration of the Foundations*, 17–21 June 2001, Växjö University Press, Växjö Sweden, (2002). This volume contains several papers relevant to the informational approach to QM. Particularly relevant are those by Fuchs and by Mermin.
2. A. Khrennikov, editor, Proceedings of the Conference, *Foundations of Probability and Physics*, 25 Nov. – 1 Dec. 2000, World Scientific, Singapore (2001).
3. L. E. Ballentine, ref. 2, pp. 71–84.
4. E. T. Jaynes, *Probability Theory: The Logic of Science*, Cambridge University Press, Cambridge, (2003).
5. L. E. Ballentine, *Quantum Mechanics: a Modern Development*, World Scientific, Singapore, (1998), sec. 1.5.
6. L. P. Hughston, R. Joza,, and W. K. Wooters, *Phys. Let. A* **183**, 14–18 (1993).
7. K. A. Kirkpatrick, *Found. Phys.* **19**, 95–102 (2006).
8. A. Einstein, B. Podolsky, and N. Rosen, *Phys. Rev.* **47**, 777–780 (1935).